\documentclass[aps,pra,twocolumn,superscriptaddress,showpacs,floatfix,]{revtex4-2}
\usepackage{float}
\usepackage{amsmath,amssymb,bm}
\usepackage{graphicx}
\usepackage{comment}
\graphicspath{{figs/}}
\usepackage{xcolor}
\usepackage{cancel}
\usepackage[colorlinks=true,linkcolor=blue,citecolor=blue,urlcolor=blue]
{hyperref}
\usepackage{orcidlink}       

\newcommand{\at}{A}
\newcommand{\tr}{t_{r}}
\newcommand{\weff}{\Delta\omega_{\rm eff}}
\newcommand{\fig}[1]{Fig.~\ref{#1}}
\newcommand{\eq}[1]{Eq.~(\ref{#1})}

\begin{document}


\title{Real-time spectral combs from delayed quantum feedback in multimode cavity QED}

\author{Diego Aranda-Lozano\,\orcidlink{0000-0003-3324-1719}}
\affiliation{Centro de Investigación en Ciencias, Universidad Autónoma del Estado de Morelos, Av.\ Universidad 1001, Col.\ Chamilpa, Cuernavaca, Morelos 62209, Mexico}
\affiliation{Instituto de Ciencias F\'isicas, Universidad Nacional Aut\'onoma de M\'exico, Avenida Universidad s/n, Col.\ Chamilpa, Cuernavaca, Morelos 62210, Mexico}
\author{Héctor M. Castro-Beltrán\,\orcidlink{0000-0002-3400-7652}}
\affiliation{Centro de Investigación en Ingeniería y Ciencias Aplicadas, Universidad Autónoma del Estado de Morelos, Avenida Universidad 1001,
62209 Cuernavaca, Morelos, Mexico}
\author{Jos\'e R\'ecamier\,\orcidlink{0000-0002-5995-0380}}
\affiliation{Instituto de Ciencias F\'isicas, Universidad Nacional Aut\'onoma de M\'exico, Avenida Universidad s/n, Col.\ Chamilpa, Cuernavaca, Morelos 62210, Mexico}
\author{Ricardo Rom\'an-Ancheyta\,\orcidlink{0000-0001-6718-8587}}
\affiliation{Centro de Física Aplicada y Tecnología Avanzada, Universidad Nacional Autónoma de México, Boulevard Juriquilla 3001,
76230 Querétaro, Mexico}
\author{Carlos A. Gonz\'alez-Guti\'errez\,\orcidlink{0000-0002-1734-1405}}
\email{carlosgg@icf.unam.mx}
\affiliation{Instituto de Ciencias F\'isicas, Universidad Nacional Aut\'onoma de M\'exico, Avenida Universidad s/n, Col.\ Chamilpa, Cuernavaca, Morelos 62210, Mexico}

\date{\today}

\begin{abstract}
We study the time-dependent physical spectrum of light for a two-level atom coupled to a large, but discrete, number of modes of an electromagnetic cavity. Radiation emitted by the atom, placed at the center of the cavity, returns after the round-trip time $\tr$, re-excites it, and is re-emitted, producing a sequence of atomic population revivals. 
For a coupling approximately flat over the atomic linewidth, the emission dynamics reduces to a delay-differential equation of Bixon--Jortner type. Using its analytical solution in terms of Laguerre polynomials, we investigate how successive emission and reabsorption cycles shape the time-dependent spectrum.
The physical spectrum exhibits revivals during which a transient frequency comb of spacing $2\pi/\tr$ develops. The comb results from the quantum interference between successive emission stages; its teeth sit not at the cavity modes but at the resonances of the effective two half-cavities created by the resonantly reflecting atom. The visibility scales as $e^{-\Gamma\tr}$ with the filter bandwidth $\Gamma$, and at long times the comb converges to the dressed-mode spectrum, which we also obtain analytically in quantitative agreement with exact diagonalization. 
We show that the physical spectrum resolves
the feedback phase, to which the atomic population
is insensitive in the regime of well-separated
emission stages, and provides additional
frequency-resolved signatures of the atomic position.
We discuss two feasible experimental implementations of our results with transmons (artificial atoms) coupled to microwave and surface-acoustic-wave resonators.
\end{abstract}

\maketitle


\section{Introduction}

The interaction between an atom and the quantum electromagnetic field is at the core of quantum optics. In free space, spontaneous emission provides its paradigmatic example: an initially excited atom irreversibly transfers its energy to the continuum of modes. However, when this emitter is placed inside an electromagnetic cavity, the boundary conditions modify the mode structure available for the interaction and can substantially alter the emission behavior~\cite{Purcell_46,Milonni1974,Cook1987}. If the cavity supports many modes within the atomic linewidth, the emitted radiation propagates, reflects from the mirrors, and returns to the position of the atom~\cite{parker1987transient}. In this regime, the non-Markovian dynamics is governed not only by the initial emission but also by the atom’s subsequent interactions with its previously emitted radiation~\cite{BuzekPRA99}, on timescales associated with the cavity round-trip time~\cite{Krimer2014}. This situation is a prototype example of \emph{delayed quantum feedback}, which has been studied for atoms in front of a mirror~\cite{Dorner2002,Tufarelli2013,
Pichler2017PNAS,Guimond2017,Vodenkova2024}, for emitters in waveguides with time-delayed coupling \cite{Pichler2016,KnuPRL2020,Vodenkova2024, Cilluffo2026}, for multimode cavity--QED \cite{Krimer2014}, driven two-level systems under feedback--loop scattering \cite{Grimsmo_PRL_2015}, and in general to open quantum systems suffering delayed coherent feedback \cite{Carmichael_QST_2017}.
Experimentally, related multimode effects have been explored using superconducting artificial atoms \cite{Sundaresan2015,Hoi2015,Mirhosseini2019,  VadirajPRA2021}, while giant-atom architectures have enabled the observation of non-Markovian dynamics and engineered light–matter interactions \cite{Andersson2019,Kannan2020,Xiao2025}.

Nevertheless, the transient spectral response associated with finite propagation delays remains less explored.

In this context a natural question arises: how does the radiation emission spectrum evolve as an atom undergoes successive emission and reabsorption cycles in a multimode cavity? Addressing this question is the central goal of this work and requires a spectral framework capable of capturing the inherently nonstationary nature of the emitted radiation. 
Because the atom's spectral response may change substantially during emission, the time-resolved physical spectrum of light, introduced by Eberly and W\'odkiewicz (EW) in~\cite{Eberly1977}, provides the operational tool needed to better analyze these processes~\cite{Eberly_1980,Eberly_PRA_1982}. 

The EW physical spectrum accounts for the finite response time of a frequency-selective detector~\cite{ficek2017quantum, sanchez1983theory}. By satisfying the time-energy uncertainty relation, it defines the spectral response accumulated up to a given observation time and thereby captures transient features that no stationary spectrum can resolve~\cite{Jia_Wang_PRA_26,Yamaguchi_PPR_22,Ancheyta_PRA_2018,Medina_2026}. In recent years, the EW spectrum has proven particularly useful for characterizing quantum beats~\cite{Castro-Ancheyta2025PRR}, nonlinear oscillators~\cite{Ancheyta_PRA_2024}, non-thermal baths~\cite{Ancheyta_2020}, strongly interacting defects in amorphous materials~\cite{Ancheyta_2022}, and quantum thermometers operating in the ultrastrong coupling regime~\cite{Ancheyta_2021}.

In this work, we apply the versatile EW formalism to a two-level atom coupled to a large but discrete number of modes of a one-dimensional multimode cavity with delayed quantum feedback. Our main findings are as follows. 

\subsection*{Overview of the results}

(i) In the flat-coupling approximation, the multimode cavity dynamics can be mapped onto a
Bixon--Jortner-type delay equation. Its analytical
solution provides a quantitative description of
the atomic population revivals and identifies
the successive emission stages responsible for
the time-dependent spectral response.


(ii) The physical spectrum evolves from the free-space spontaneous-emission profile to a sequence of spectral revivals, during which a transient comb structure with spacing $2\pi/\tr$ appears. Our analytic solution clearly identifies this comb as the quantum interference between radiation emitted in successive stages of the dynamics. The comb teeth sit, exactly, at the resonances of two half-cavities that are created by the atom, which near resonance acts as a mirror, and gives its visibility in terms of the filter bandwidth. 

(iii) At long times and for a narrow filter, the transient comb converges to the discrete dressed-mode spectrum of the atom--cavity system, whose frequencies we obtain in closed form from the poles of the Laplace-transformed amplitude, in quantitative agreement with the exact eigenfrequencies of the multimode model. 

(iv) The feedback phase (the phase acquired by the emitted field in one round trip) 
and the atomic position are resolved by the EW spectrum, whereas the excited-state population, in the regime considered, is insensitive to this feedback phase.
The EW spectrum therefore carries valuable information that the atomic population dynamics does not. 

(v) We discuss feasible superconducting circuit implementations considering a transmon coupled to a microwave transmission line or to a surface-acoustic-wave resonator.

The paper is organized as follows. Section~\ref{sec:model} introduces the atom-cavity model and presents the numerical and the exact solutions of the delay-differential equation. Section~\ref{sec:spectrum} introduces the time-dependent physical spectrum, presents the numerical spectra, and derives their analytic structure, including the comb and the crucial role of the filter bandwidth. Section~\ref{sec:control} shows how the comb is controlled by the feedback phase and the atomic position, and Sec.~\ref{sec:longtime} discusses the long-time limit. Section~\ref{sec:cqed} outlines a possible experimental implementation, and Sec.~\ref{sec:conclusions} summarizes our results. Technical details with mathematical derivations are given in the appendices.

\section{Multimode cavity-QED model}
\label{sec:model}

\subsection{System Hamiltonian}
We consider a two-level atom interacting with a multimode one-dimensional cavity of length $L$ within the dipole and the rotating-wave approximation (RWA). The Hamiltonian is~\cite{BuzekPRA99}
\begin{equation}
H=\hbar\omega_a\sigma_+^{}\sigma_{-}^{}\!+\!\sum_{n=1}^{N}\!\hbar\omega_n^{} a_n^\dagger a_n^{}
\!-\!\sum_{n=1}^{N}\!\hbar g_n(x)(a_n^{}\sigma_+^{}\!+\!a_n^\dagger\sigma_-^{}),
\label{eq:H}
\end{equation}
where $\omega_n=n\pi c/L$ and $a_n$ ($a_n^\dagger$) are the frequency and annihilation (creation) operator of the $n$th mode, $c$ is the speed of light, $\omega_a$ is the atomic transition frequency, and $\sigma_\pm^{}$ are the atomic raising and lowering operators. The zero of energy is chosen at the atomic ground state and the field in the vacuum state. 
The position-dependent coupling is
\begin{equation}
g_n(x)=g_0^{}\sqrt{\frac{\omega_n}{\omega_a}}\sin(k_n x),
\qquad k_n=\frac{\omega_n}{c}=\frac{n\pi}{L},
\label{eq:gn}
\end{equation}
with $g_0^{}=d_{eg}\sqrt{\omega_a/\hbar\epsilon_0L}$, being $d_{eg}$ the transition dipole moment.
A  full microscopic derivation of \eq{eq:H} for a transmon (artificial atom) coupled to a multimode transmission-line resonator is given in Appendix~\ref{sec:circuit_qed}; the physical discussion can be found in Sec.~\ref{sec:cqed}.

Since the Hamiltonian $H$ conserves the total number of excitations, an initially excited atom with the field in the vacuum remains in the single-excitation manifold
\begin{equation}
|\psi(t)\rangle=\alpha(t)\,|e,\vec 0\rangle+\sum^{N}_{n=1}\beta_n(t)\,|g,1_n\rangle,
\label{eq:state}
\end{equation}
where $|\vec 0\rangle$ is the multimode vacuum state and $|1_n\rangle\!\equiv\! a_n^\dagger|\vec 0\rangle$. The Schr\"odinger equation  $i\hbar\partial_t|\psi(t)\rangle\!=H\!|\psi(t)\rangle$ yields
\begin{subequations}\label{coupled_eqs}
\begin{align}
i\dot\alpha(t)&=\omega_a\,\alpha(t)-\sum_{n=1}^{N} g_n(x)\,\beta_n(t),
\label{eq:alpha}\\
i\dot\beta_n(t)&=\omega_n\,\beta_n(t)-g_n(x)\,\alpha(t),
\label{eq:beta}
\end{align}
\end{subequations}
with $\alpha(0)=1$, $\beta_n(0)=0$.
The system of Eqs.~(\ref{coupled_eqs}) will be integrated numerically and compared with the analytical solution presented in Sec.~\ref{sec:dde}. The excited-state population is $P_e(t)=|\alpha(t)|^2$. With the energy zero at the atomic ground state, the phase of $\alpha(t)$ rotates at the atomic transition frequency, since $|e,\vec0\rangle$ has energy $\hbar\omega_a$ and $|g,\vec0\rangle$ has energy zero. Because the detected field is the dipole far field, $E^{(+)}\propto\sigma_-$, whose correlation function is $\alpha^*(t_1)\alpha(t_2)$ (Sec.~\ref{sec:definition}), this rotation is the optical carrier of the emitted radiation. The slowly varying envelope of $\alpha(t)$ is introduced in Sec.~\ref{sec:dde}, once the radiative correction to the carrier frequency has been obtained.


\subsection{Numerical results of the light-matter dynamics}
\label{sec:numerics}

Throughout this work the atom is initially in its excited state and located at the center of the cavity, $x=L/2$, unless stated otherwise, and the field is initially in the vacuum. We use the same parameters and dimensionless units as in~\cite{BuzekPRA99} with $\hbar=c=1$ and $L=2\pi$, so that the fundamental mode frequency is $\omega_1=1/2$. At the cavity center only the odd modes couple to the atom, since $\sin(k_nL/2)=\sin(n\pi/2)=0$ for even $n$; the spacing between consecutive coupled modes is therefore
$\weff=\omega_{n+2}-\omega_n=1 \quad (n\ \text{odd}).$
The radiation emitted by the atom is split into two and travels to the cavity boundaries on either side, a distance $L/2$ away, and returns to the atomic position after the round-trip time
$\tr=L/c=2\pi.$
The parameters used in our simulations are $g_0^2=1/2$, frequency cutoff $\omega_{\rm cut}=200$, and $\omega_a=\omega_{\rm cut}/2$, corresponding to $N=400$ cavity modes. In all figures the time is measured in units of the round-trip time, $t/\tr$, and frequencies as $(\omega-\tilde\omega_a)/\weff$, with $\tilde\omega_a$ the Lamb-shifted atomic frequency introduced below Eq.~(\ref{flat_kernel}). 
Note that the atomic frequency $\omega_a=100$ corresponds to $n=200$, an \emph{even} mode that does not couple to a centered atom and the atom sits exactly halfway between the coupled modes at $\omega=99.5$ and $100.5$. The crucial consequences of this choice are discussed in Secs.~\ref{sec:combstructure} and \ref{sec:control}.


\begin{figure*}[t]
\includegraphics[width=0.95\textwidth]{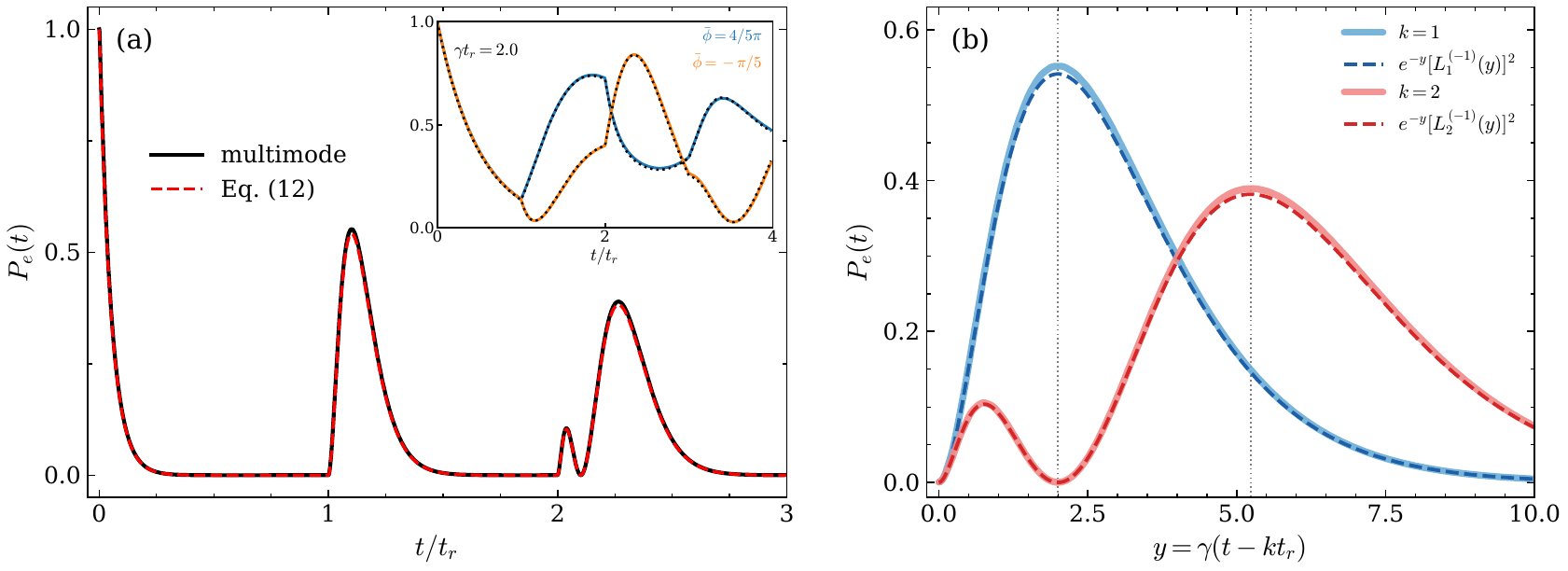}
\caption{ 
(a) Excited-state atomic population  
$P_e(t)=|\alpha(t)|^2=|\at(t)|^2$ versus $t/\tr$ for an atom at the cavity center, initially excited, with the field in the vacuum, from the full multimode numerical simulation (solid) and from the exact analytical solution (dashed), Eq.~(\ref{eq:solution}), of the delay-differential equation. Parameters: $L=2\pi$, $c=1$, $g_0^2=1/2$, $\omega_a=100$, $\omega_{\rm cut}=200$ ($N=400$ modes, $\omega_1=0.5$), giving $\gamma\tr=19.7$; these parameters are used in all figures unless stated otherwise. Inset: same comparison for $g_0^2=0.05$ ($\gamma\tr=2.0$, $\Delta_L=-0.1$) and two feedback phases, $\omega_a=100$ ($\bar\phi=4\pi/5$) and $\omega_a=100.5$ ($\bar\phi=-\pi/5$); dotted lines are the analytic solution. When the stages overlap the population becomes phase sensitive. (b) The first ($k=1$) and second ($k=2$) revivals of the numerical simulation plotted against $y=\gamma(t-k\tr)$ (solid lines), compared with the universal forms $e^{-y}[L_k^{(-1)}(y)]^2$ of \eq{eq:Pe_universal} (dashed). Dotted lines mark $y=2$ and $y=3+\sqrt5$.}
\label{fig:compare}
\end{figure*}

Figure~\ref{fig:compare}(a) shows the atomic excited-state population as a function of $t/t_r$. Despite being located inside a cavity that supports a large but finite number of modes, the atomic initial decay is exponential and indistinguishable from free-space decay; this regime ends on a time scale shorter than $\tr$. At $t\sim\tr$ the two counter-propagating wave packets reflected by the cavity mirrors return simultaneously, interfere constructively at the atom central position, and partially re-excite it, producing the first revival. A second revival follows the next round trip, at $t\simeq 2\tr$, preceded by a smaller precursor. Although the revivals are not instantaneous, the population rises sharply from zero after each round trip and peaks about $1/\gamma$ later, with $\gamma\equiv2\pi g_0^2/\weff$ being the effective free-space decay rate. The analytic solution presented in the following subsection reproduces and expands these features quantitatively.

\subsection{Delay-differential equation and exact solution}
\label{sec:dde}

In this subsection, we show that the atomic population dynamics of the multimode cavity-QED Hamiltonian given by Eq.~(\ref{eq:H}) can be solved exactly in the flat coupling limit. The analytical approach adopted here is inspired by the methods used in~\cite{barnett2002methods,cevheroglu2026} to solve the Bixon–Jortner model, whose equations share a similar structure with those governing the dynamics of our system. We outline the main and nontrivial algebraic steps. By eliminating the field amplitudes $\beta_n(t)$ from Eqs.~(\ref{coupled_eqs}), one obtains the following exact integro-differential equation
\begin{equation}\label{eq:integro}
\dot{\alpha}(t)=-i\omega_a\alpha(t)-\int_0^t d\tau\, K(\tau)\,\alpha(t-\tau),
\end{equation}
with the kernel $K(\tau)=\sum_n g_n^2\,e^{-i\omega_n\tau}$.
For a centered atom ($x=L/2$) the sum runs over the odd modes, $\omega_n=(m+\tfrac12)\weff$, and the coupling (\ref{eq:gn}) can be written identically as $g_n^2=g_0^2[1+(\omega_n-\omega_a)/\omega_a]$. For $\omega_a\gg\{\gamma,\weff\}$ the flat term dominates near resonance, while the linear term produces a frequency-independent shift of the atomic line (see the Appendix~\ref{app:dde})
\begin{equation}
\Delta_L=\sum_n\frac{g_n^2}{\omega_a-\omega_n}\simeq-\frac{g_0^2 N_c}{\omega_a} = -1,
\label{eq:lamb}
\end{equation}
for our parameters.
This is the Lamb shift produced by the $\sqrt{\omega_n}$ dependence of the coupling and by the cutoff; as usual in rotating-wave models it is cutoff dependent. Separating this constant from the kernel, \eq{eq:integro} reads
\begin{subequations}
\label{eq:integro_flat}
\begin{align}
\dot{\alpha}(t)&=-i\tilde\omega_a\,\alpha(t)-\int_0^t d\tau\, K_{\rm flat}(\tau)\,\alpha(t-\tau),
\\
K_{\rm flat}(\tau)&=g_0^2\sum_{n\ {\rm odd}} e^{-i\omega_n\tau},
\label{flat_kernel}
\end{align}
\end{subequations}
with $\tilde\omega_a=\omega_a+\Delta_L$. The only effect of the linear part of the coupling is to renormalize the atomic frequency, and what remains is the flat kernel. Now that the carrier frequency is known, we factor it out and work with the envelope
\begin{equation}
\at(t)\equiv e^{i\tilde\omega_a t}\,\alpha(t),
\label{eq:envelope}
\end{equation}
which varies only on the time scales of the atomic emission dynamics, with $P_e=|\at(t)|^2$. In terms of $\at(t)$, Eqs.~(\ref{eq:integro_flat}) become
\begin{subequations}
\begin{align}
\dot{\at}(t)&=-\int_0^t d\tau\, \bar K(\tau)\,\at(t-\tau),
\\
\bar K(\tau)&=e^{i\tilde\omega_a\tau}K_{\rm flat}(\tau)\simeq\gamma\sum_{k=0}^{\infty}e^{ik\bar\phi}\,\delta(\tau-k\tr),
\label{eq:kernel}
\end{align}
\end{subequations}
where the Dirac comb follows from the Poisson summation formula (see the Appendix~\ref{app:dde}), and the delta function at $\tau=0$, which sits at the boundary of the integration range, contributes with weight $1/2$. Here
\begin{equation}
\bar\phi\equiv\tilde\omega_a\tr+\pi \pmod{2\pi}
\label{eq:phibar}
\end{equation}
is the \emph{feedback phase}, the phase accumulated by the field at the (renormalized) atomic frequency during one round trip, plus the $\pi$ phase shift upon reflection at a mirror. For the parameters of \fig{fig:compare}(a), $\Delta_L\tr=-2\pi$, so $\bar\phi=\omega_a\tr+\pi=\pi$; this means that $\tilde\omega_a$ lies halfway between two coupled modes. The corrections to \eq{eq:kernel} are of relative order $\gamma/\omega_a$, $\weff/\omega_a$, and $\Delta_L/\omega_a$. Equation~(\ref{eq:kernel}) is then a delay-differential equation
\begin{equation}
\dot{\at}(t)=-\frac{\gamma}{2}\,\at(t)-\gamma\sum_{k\geq1}e^{ik\bar\phi}\,\at(t-k\tr)\,\Theta(t-k\tr),
\label{eq:dde}
\end{equation}
which is the two-mirror analog of the half-cavity equation of Dorner and Zoller \cite{Dorner2002}. The first term is free-space spontaneous emission with decay rate $\gamma$; for the parameters of \fig{fig:compare}(a), $\gamma=2\pi g_0^2/\weff=\pi$. The $k$th feedback term describes the return of radiation emitted $k$ round trips earlier; in a cavity, unlike in front of a single mirror, all multiples of $\tr$ appear because the field keeps bouncing between the mirrors. Note that the feedback phase involves $\tilde\omega_a$, not $\omega_a$. The atom feeds back on itself with the phase set by its renormalized frequency.

Equation~(\ref{eq:dde}) is mathematically equivalent to the delay-differential equation governing the Bixon--Jortner
model, whose analytical solution can be expressed in terms of Laguerre polynomials \cite{barnett2002methods,cevheroglu2026}. Here we adapt this result to the multimode cavity-QED
system, accounting for the feedback phase and the renormalized atomic frequency.
For completeness, we provide a derivation based on the Laplace transform in Appendix~\ref{app:dde}.
Defining $s_k^{}\!=\!t\!-\!k\tr$ and $y_k^{}\!=\!\gamma s_k^{}$, the exact solution of Eq. (\ref{eq:dde}) reads,
\begin{equation}
\at(t)=\sum_{k=0}^{\lfloor t/\tr\rfloor}e^{ik\bar\phi}\,e^{-y_k/2}\left[L_k(y_k^{})-L_{k-1}(y_k^{})\right],
\label{eq:solution}
\end{equation}
with $L_k$ the Laguerre polynomials and $L_{-1}\equiv0$. This solution consists of a sequence of delayed emission stages. Using $L_k-L_{k-1}=L_k^{(-1)}=-(y/k)L_{k-1}^{(1)}$, the first stages read
\begin{subequations}
\begin{align}
k=0:&\quad e^{-\gamma t/2},\label{free_evolution}\\
k=1:&\quad -e^{i\bar\phi}\,\gamma s_1\,e^{-\gamma s_1/2},\\
k=2:&\quad e^{2i\bar\phi}\,\gamma s_2\Big(\frac{\gamma s_2}{2}-1\Big)e^{-\gamma s_2/2}.
\label{eq:terms}
\end{align}
\end{subequations}
Each stage with $k\ge1$ vanishes at $s_k^{}=0$: re-excitation by the returning pulse starts from zero amplitude and builds up over a time $\sim1/\gamma$, exactly as seen in \fig{fig:compare}.

In the regime $\gamma\tr\gg1$ the pulses do not overlap (unlike the single-mode Jaynes-Cummings model~\cite{Castro_PRA_96}),
so on the $k$th interval only the $k$th term survives and
\begin{equation}
P_e(t)\simeq e^{-y_k}\left[L_k(y_k)-L_{k-1}(y_k)\right]^2,\quad k\tr\le t<(k+1)\tr.
\label{eq:Pe_universal}
\end{equation}
When expressed in the scaled time $y_k^{}=\gamma(t-kt_r)$, Eq.~\eqref{eq:Pe_universal} contains no parameter at all. The shape of each revival is the same for any $\gamma$, $t_r$ and $\bar\phi$, which enter only through the scale $1/\gamma$ and the delay $kt_r$. Note that Eq.~(\ref{eq:Pe_universal}) predicts: (i)~a first revival of height $4/e^2\simeq0.541$ at $t=\tr+2/\gamma$; (ii)~a second revival with a precursor of height $0.104$ at $y_2=3-\sqrt5$, an exact zero at $y_2=2$, and a main peak of height $0.382$ at $y_2=3+\sqrt5$; (iii)~for the parameters of \fig{fig:compare}(a) ($\gamma\tr=19.7$), peaks at $t/\tr\simeq1.10$, $2.04$, and $2.27$ and a zero at $t/\tr\simeq2.10$. 
Figure~\ref{fig:compare}(a) compares the complete analytical solution (\ref{eq:solution}) with the full multimode numerical simulation; the two are almost indistinguishable on the scale of the figure. 
Figure~\ref{fig:compare}(b) shows the first and second revivals plotted against $y=\gamma(t-k\tr)$, these almost collapse onto the parameter-free curves $e^{-y}[L_k^{(-1)}(y)]^2$. The inset of \fig{fig:compare}(a) shows that for $\gamma\tr\simeq2$, where the stages overlap and interfere, the agreement persists only if the feedback phase is evaluated at the renormalized frequency, $\bar\phi=\tilde\omega_a\tr+\pi$, rather than at $\omega_a$. The approximate $2\%$ residual differences in the peak heights arise from the $\sqrt{\omega_n}$ dependence of the coupling and from the finite cutoff, both neglected in \eq{eq:kernel}.
The fact that the population revivals are insensitive to the feedback phase $\bar\phi$ in the regime $\gamma t_r\gg 1$ (the phase appears only in cross terms between different $k$, which are suppressed by $e^{-\gamma\tr/2}$) will be contrasted below with the behavior of the time-resolved spectrum, which resolves this feedback phase.

\section{Time-resolved physical spectrum}
\label{sec:spectrum}
\subsection{The Eberly-W\'odkiewicz definition}
\label{sec:definition}
Since the global atom--field evolution in the lossless multimode cavity is unitary (nonstationary), conventional spectroscopy based on the time-independent (stationary) Wiener-Khintchine emission spectrum can not be applied directly. An operational alternative is the EW time-dependent physical spectrum~\cite{Eberly1977}, which describes the counting rate of a photodetector preceded by a frequency-selective filter of finite  bandwidth $\Gamma$. 
The EW spectrum is defined as
\begin{equation}\label{eq:S_def}
\begin{aligned}
	{S}(\omega,\!\Gamma,\!t)\!=\! 
	2\Gamma e^{-2\Gamma t}\!\! \int_{0}^{t}\! dt_{1}^{}\!\!\int_{0}^{t}\!\!dt_{2}^{} e^{(\Gamma - \mathrm{i}\omega)t_{1}} e^{(\Gamma + \mathrm{i}\omega)t_{2}}G(t_{1},\!t_{2}),
\end{aligned}
\end{equation}
where $G(t_1,t_2)=\langle E^{(-)}(t_1)E^{(+)}(t_2)\rangle$ is the first-order correlation function of the field entering the filter. 
The detected field originates from the atomic dipole, so $E^{(\pm)}(t) \propto \sigma_{\mp}^{}(t)$ 
and $G(t_1,t_2)=\langle\sigma_+(t_1)\sigma_-(t_2)\rangle$.
For the initial condition considered in Sec.~\ref{sec:model} the correlation reduces to (see Appendix~\ref{app:corr})
\begin{equation}
G(t_1,t_2)=\alpha^*(t_1)\alpha(t_2)=e^{-i\tilde\omega_a(t_2-t_1)}\,\at^*(t_1)\at(t_2),
\label{eq:G_alpha}
\end{equation}
and \eq{eq:S_def} factorizes into a modulus squared~\cite{Ancheyta_PRA_2024}
\begin{align}
S(\omega,\Gamma,t)
=2\Gamma\,e^{-2\Gamma t}\left|\int_0^t dt'\,e^{(\Gamma+i\delta)t'}\,\at(t')\right|^2,
\label{eq:S_alpha}
\end{align}
with $\delta\equiv\omega-\tilde\omega_a$.
Equation~(\ref{eq:S_alpha}) is the central working formula in this work. The physical spectrum at time $t$ is the modulus squared of a finite-time, exponentially weighted Fourier transform of the atomic amplitude. In terms of the envelope, the frequency variable is naturally the detuning $\delta$ from the renormalized line $\tilde\omega_a$, which we use throughout.

\begin{figure*}[t]
    \centering
    \includegraphics[width=0.98\textwidth]{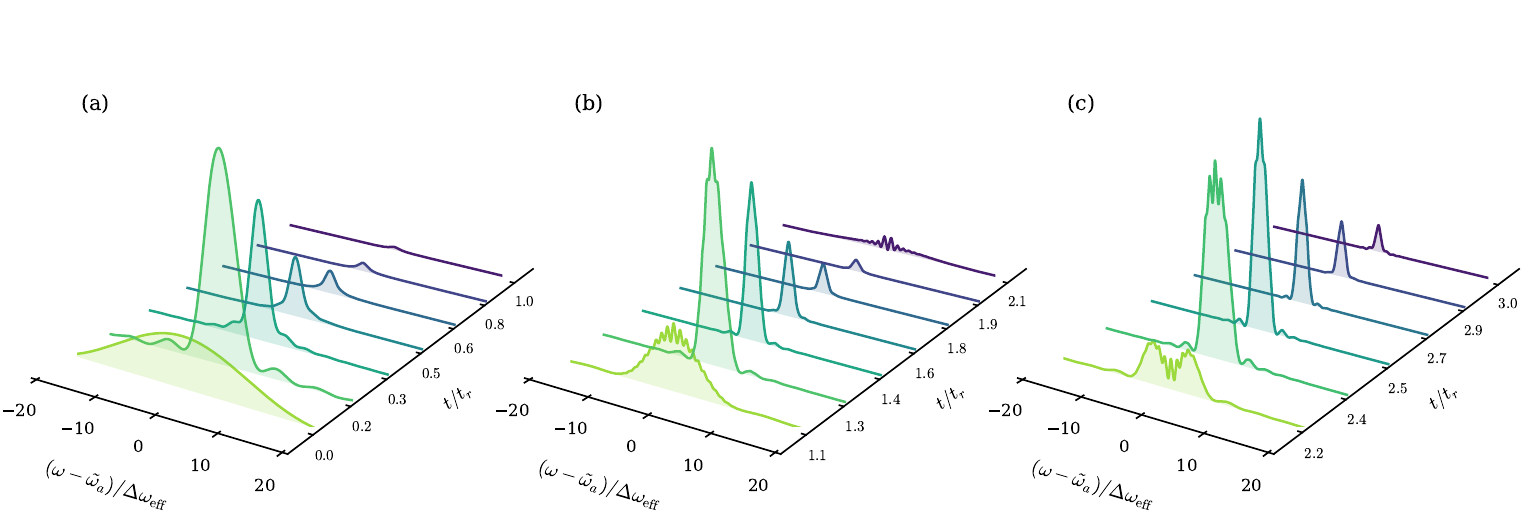}

    \vspace{0.2cm}

    \includegraphics[width=0.98\textwidth]{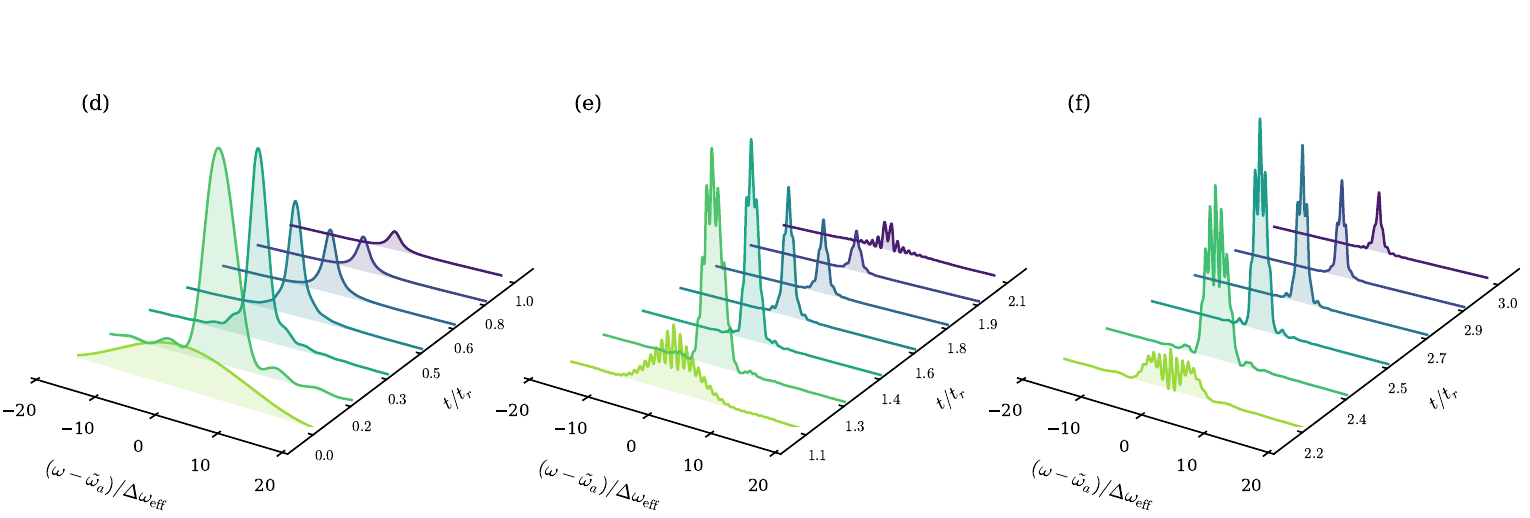}

    \caption{
    Time-dependent physical spectrum $S(\omega,\Gamma,t)$ for an atom at the center of a multimode electromagnetic cavity, over three successive round-trip intervals: (a,d) initial spontaneous emission, $t/\tr\leq 1$; (b,e) first spectral revival, $1\leq t/\tr\leq 2$; (c,f) second revival, $2\leq t/\tr\leq 3$. 
    The upper and lower panels correspond to filter bandwidths $\Gamma\tr=\pi$ ($\Gamma/\gamma=0.16$) and $\Gamma\tr=1.9$ ($\Gamma=0.3$, $\Gamma/\gamma=0.095$), respectively. The narrower filter increases the comb visibility during the first revival from $\simeq 0.03$ to $\simeq 0.15$. We use the same parameters as in \fig{fig:compare}.}
    \label{fig:spectrum}
\end{figure*}

During the effective free-space decay the atomic amplitude is well approximated by $\at(t)=e^{-\gamma t/2}$, and \eq{eq:S_alpha} produces a birth-to-death evolution of the spectrum [see \fig{fig:spectrum}(a)] with the closed-form expression given by \cite{Castro-Ancheyta2025PRR}
\begin{equation}
S_{\rm free}^{}=2\Gamma\,
\frac{e^{-\gamma t}+e^{-2\Gamma t}-2e^{-(\Gamma+\gamma/2)t}\cos(\delta t)}
{(\Gamma-\gamma/2)^2+\delta^2}.
\label{eq:S_free}
\end{equation}
$S_{\rm free}^{}$ has a $\mathrm{sinc}^2$-like profile of width $\sim 2\pi/t$ at short times and tends, for $\Gamma<\gamma/2$, to a Lorentzian of half-width $\gamma/2-\Gamma$ times the decaying prefactor $e^{-2\Gamma t}$.
\subsection{Numerical results for the spectral evolution}
\label{sec:specnumerics}

Figure~\ref{fig:spectrum} shows spectral traces of the EW spectrum computed from the amplitude $\at(t)$ over three successive emission cycles, for $\Gamma\tr=\pi$ ($\Gamma=0.5$). During the first stage [\fig{fig:spectrum}(a)] the spontaneously emitted radiation has not yet reached the mirrors, so the atom cannot distinguish the cavity from free space, and the numerical spectrum coincides with \eq{eq:S_free}. At short times ($t/\tr\lesssim0.1$) the spectrum is broad, because the short observation time limits the frequency resolution. As time increases, it narrows and develops the $\mathrm{sinc}^2$-like profile centered at $\tilde\omega_a$. Once the emitted light pulse has passed through the filter, no further radiation reaches the detector and the spectrum dies as $e^{-2\Gamma t}$.

After one round trip, the returning field radiation re-excites the atom and induces renewed emission, which manifests as the first spectral revival [\fig{fig:spectrum}(b)]. Around $t/\tr\approx1.1$ a transient comb-like modulation appears on the spectral envelope. At $t/\tr\approx1.3$ the spectrum reaches its maximum, with a dominant central peak and side shoulders. The second revival [\fig{fig:spectrum}(c)] follows the same mechanism. Around $t/\tr\approx2.1$--$2.2$ the modulation is deeper and the peaks sharper; near the maximum at $t/\tr\approx2.5$ a narrow central peak dominates, after which the amplitude decays again. Two systematic trends are visible. First, successive revivals have narrower and more intense peaks. This follows from \eq{eq:terms}: the $k$th emission stage has the temporal profile $|L_k^{(-1)}(\gamma s)|e^{-\gamma s/2}$, which extends over a longer time for larger $k$ ($s\,e^{-\gamma s/2}$ for $k=1$, $s(\gamma s/2-1)e^{-\gamma s/2}$ for $k=2$), and a longer emission profile produces a narrower spectral line. Second, the comb modulation becomes more pronounced from one revival to the next. Both trends are explained quantitatively in Sec.~\ref{sec:combstructure}.

\begin{figure}[t]
\includegraphics[width=\columnwidth]{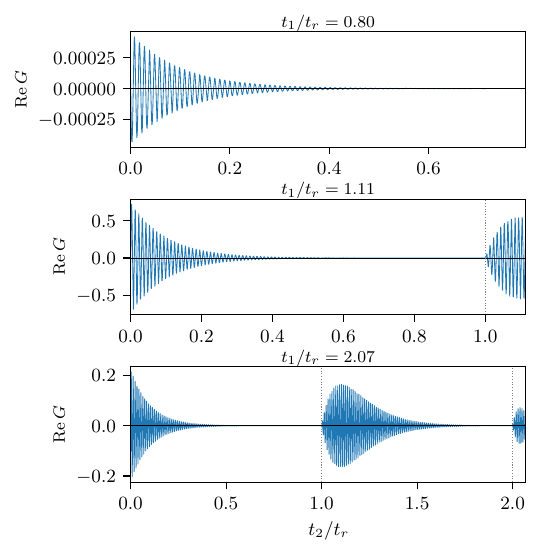}
\caption{Real part of the correlation function $G(t_1,t_2)$, \eq{eq:G_alpha}, as a function of $t_2/\tr$ at fixed $t_1/\tr=0.80$ (top), $1.11$ (center), and $2.07$ (bottom). The fast oscillation is due to the optical carrier $e^{-i\tilde\omega_at_2}$. Successive emission stages separated by $\tr$ are visible as distinct bursts; the comb in the spectrum arises from the interference between them.}
\label{fig:corr}
\end{figure}

We trace the origin of the modulation back to the correlation function itself. Figure~\ref{fig:corr} shows cuts of $\mathrm{Re}[G(t_1,t_2)]$ at $t_1/\tr=0.80$, $1.11$, and $2.07$. At $t_1/\tr=0.80$ only the initial emission contributes. At $t_1/\tr=1.11$ the initial burst and the first re-emission appear as two separate oscillatory bursts separated by $\simeq\tr$; at $t_1/\tr=2.07$ three bursts are present. At fixed $t_1$ the cut $G(t_1,t_2)=\alpha^*(t_1)\alpha(t_2)$ is, up to a constant, the amplitude $\alpha(t_2)$ itself, so \fig{fig:corr} displays directly the sequence of emission bursts that enters \eq{eq:S_alpha}. The spectrum is the squared modulus of the Fourier transform of this sequence. Two bursts separated by $\tr$ transform into the single-burst envelope multiplied by $|1+\eta\,e^{i\delta\tr}|^2$, with $\eta$ their relative amplitude, which oscillates in $\delta$ with period $2\pi/\tr$. As in two-pulse interference, the fringe spacing is the inverse of the delay between the pulses \cite{Lepetit:95}. This is the comb, which in \eq{eq:S_comb} below we write out for all stages.

\subsection{Analytic structure: interference and comb}
\label{sec:combstructure}

Inserting the exact solution (\ref{eq:solution}) into \eq{eq:S_alpha} and shifting the integration variable in each term, the EW spectrum takes the form
\begin{equation}
S(\omega,\Gamma,t)=2\Gamma e^{-2\Gamma t}\left|\sum_{k=0}^{\lfloor t/\tr\rfloor}
e^{ik(\bar\phi+\delta\tr)}\,e^{k\Gamma\tr}\,F_k(\delta;t-k\tr)\right|^2,
\label{eq:S_comb}
\end{equation}
with the single-stage amplitudes
\begin{equation}
F_k(\delta;T)=\int_0^T ds\,e^{(\Gamma+i\delta-\gamma/2)s}\left[L_k(\gamma s)-L_{k-1}(\gamma s)\right].
\label{eq:Fk}
\end{equation}
Note that Eq.~(\ref{eq:S_comb}) makes the interference mechanism explicit. Each emission stage $k$ contributes an envelope $|F_k|^2$ centered at $\delta=0$ and of width $\sim\gamma$; the stages interfere through the phase factors $e^{ik(\bar\phi+\delta\tr)}$, which are periodic in $\delta$ with period $2\pi/\tr=\weff$. This is the comb. During the first revival the interference is between $k=0$ and $k=1$, and the cross term is proportional to ${\rm Re}[e^{i(\bar\phi+\delta\tr)}F_1F_0^*]$. Since $L_1^{(-1)}(y)=-y$, the first re-emission has the opposite sign to the initial emission, $F_1F_0^*\simeq-|F_1F_0|$ near $\delta=0$; with $\bar\phi+\delta\tr=\omega\tr+\pi$ this gives a modulation $\propto\cos(\omega\tr)$, so the comb maxima sit at
\begin{equation}
\omega_m\tr=2\pi m,\qquad m\in\mathbb Z ,
\label{eq:comb_positions}
\end{equation}
independently of $\omega_a$. These are \emph{not} the coupled cavity modes, $\omega\tr=(2m+1)\pi$, but the frequencies halfway between them: for our parameters, $\omega_m=98,99,100,\dots$, in agreement with the comb of \fig{fig:spectrum}(b) and with the cuts of \fig{fig:phase}. The condition $\omega\tr=2\pi m$ is the resonance condition of a cavity of length $L/2$. Its appearance has a clear physical origin: a resonant two-level atom in one dimension acts as a perfect reflector \cite{Shen2005}, and in the regime $\gamma\gg\weff$ the re-excited atom reflects the returning radiation with a $\pi$ phase shift. The atom thus effectively divides the cavity into two half-cavities, and the fields that survive repeated reflections between a mirror and the atom are those at the half-cavity resonances. The same sign alternation holds between any two consecutive stages, since $L_k^{(-1)}(y)\sim(-y)^k/k!$ at large $y$, so later revivals reinforce the same comb. Within the atomic linewidth the teeth are additionally pulled by the atom, as described by the dressed-mode equation of Sec.~\ref{sec:longtime}, and far from $\tilde\omega_a$ (where the atom is transparent) they revert to the bare coupled modes.
The relative weight of the stages is governed by the factors $e^{k\Gamma\tr}$: in \eq{eq:S_comb} the most recent stage dominates, and the stage emitted one round trip earlier is suppressed by $e^{-\Gamma\tr}$, because the filter has been forgetting it for a time $\tr$. Keeping the two most recent stages, the spectrum during the $k$th revival is a two-beam interference pattern whose Michelson contrast is \cite{Mandel_Wolf_1995} (Appendix~\ref{app:visibility})
\begin{equation}
V_k=\frac{2|c_k|}{1+|c_k|^2},\qquad c_k=e^{-\Gamma\tr}\,\frac{F_{k-1}(\delta)}{F_k(\delta)},
\label{eq:visibility}
\end{equation}
so that $V_k\simeq2e^{-\Gamma\tr}|F_{k-1}/F_k|$ when the earlier stage is weak. 
The ratio of single-stage amplitudes is of order unity and increases slowly with $k$ (the $(k-1)$th stage has had longer to be collected by the filter than the $k$th), which explains why the modulation deepens from the first to the second revival in \fig{fig:spectrum}. For the first revival and $t-\tr\gg1/\gamma$ the ratio has the closed form $|F_0/F_1|\to\tfrac12-\Gamma/\gamma$, giving $V_1\simeq e^{-\Gamma\tr}(1-2\Gamma/\gamma)$ for small contrast. 
Figure~\ref{fig:visibility} tests this prediction. For each filter bandwidth the visibility of the comb is measured on the spectrum at the maximum of the first revival ($t/\tr=1.27$), computed both from the multimode amplitude and from the analytic amplitude (\ref{eq:solution}), which agree to three digits, and compared with \eq{eq:visibility} evaluated at $\delta=0$ in two ways: with the single-stage amplitudes $F_0$ and $F_1$ collected up to the observation time, which is the exact two-stage prediction for that measurement, and with their long-time limits, which give the closed form $|c_1|=e^{-\Gamma\tr}(\tfrac12-\Gamma/\gamma)$. The measured contrast follows the $e^{-\Gamma\tr}$ dependence of \eq{eq:visibility} over a factor of 20, from $V\simeq0.65$ at $\Gamma\tr=0.5$ to $0.03$ at $\Gamma\tr=\pi$, and lies between the two evaluations, within $\sim20\%$ of either for $\Gamma\tr\lesssim2$. The residual spread is the error of treating $c_k$ as constant between a tooth and the neighboring minimum, half a tooth spacing apart, which \eq{eq:visibility} assumes (Appendix~\ref{app:visibility}). Beyond $\Gamma\tr\simeq\pi$, i.e.\ for a filter wider than half the tooth spacing, adjacent teeth are no longer resolved and the comb disappears.

The same long-time limits of \eq{eq:Fk} fix the widths of the envelopes on which the teeth sit. For $\Gamma<\gamma/2$ and $p=\gamma/2-\Gamma-i\delta$ one finds $F_0\to1/p$ and $F_1\to-\gamma/p^2$: the initial emission has a Lorentzian envelope of half-width $\gamma/2-\Gamma$, whereas the first revival has a squared-Lorentzian envelope, about $35\%$ narrower, because the re-excited atom emits over a longer time. 
Each further round trip adds another power of $1/p$, so higher-order stages are narrower still, which is the progressive sharpening of the central feature from the first to the second revival in \fig{fig:spectrum}.
\subsection{Role of the filter bandwidth}
\label{sec:filter}

\begin{figure}[t]
\includegraphics[width=0.9\columnwidth]{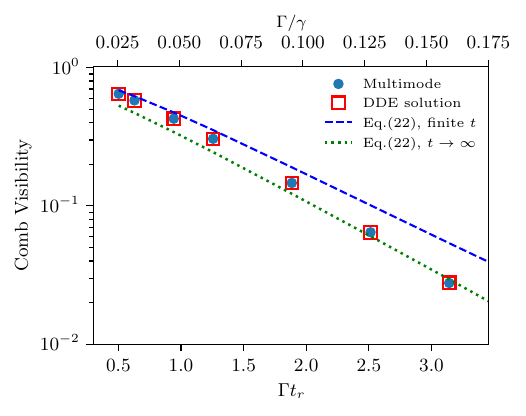}
\caption{Comb visibility during the first revival ($t/\tr=1.27$) as a function of $\Gamma\tr$ (bottom axis) and $\Gamma/\gamma$ (top axis). The values are extracted from the multimode spectra (circles) and from the spectra computed with the analytic amplitude (\ref{eq:solution}) (red squares), compared with \eq{eq:visibility} at $\delta=0$ using the finite-time amplitudes $F_{0,1}(0;t-k\tr)$ (blue dashed), and its long-time limit $|c_1|=e^{-\Gamma\tr}(\tfrac12-\Gamma/\gamma)$ (green dotted). The measured visibility is the mean of $(S_{\max}-S_{\min})/(S_{\max}+S_{\min})$ over adjacent extrema with $|\omega-\tilde\omega_a|<1.5\,\weff$ (Appendix~\ref{app:visibility}); for $\Gamma\tr\ge4.4$ no extrema remain.}
\label{fig:visibility}
\end{figure}

Equation~(\ref{eq:visibility}) identifies the parameter that controls whether the comb is observable: $\Gamma\tr=2\pi\Gamma/\weff$, since $|c_k|\propto e^{-\Gamma\tr}$. The physical content is the memory of the detector. The filter retains the field that entered it for a time $1/\Gamma$, so when the $k$th emission stage arrives the previous one has decayed to $e^{-\Gamma\tr}$ of its amplitude, and that remnant is the second beam of the interference. For $\Gamma\tr\gg1$ the filter has forgotten each stage before the next one arrives; the spectrum is then a sequence of isolated single-stage envelopes, and the discrete structure of the cavity is invisible. For $\Gamma\tr\lesssim1$, i.e., $\Gamma\lesssim\weff/2\pi$, consecutive stages coexist in the filter and the comb appears with visibility of order unity. The same condition can be read in the frequency domain: the filter resolves the comb only if its bandwidth is smaller than the tooth spacing, and for $\Gamma\gtrsim\weff/2$, i.e.\ $\Gamma\tr\gtrsim\pi$, adjacent teeth merge and no comb survives, which is where the data of \fig{fig:visibility} end. The figures of this work span this range: for $\Gamma\tr=\pi$ (\fig{fig:spectrum}) the measured visibility at $t/\tr=1.27$ is $V\sim03$, consistent with the faint modulation seen there; for $\Gamma\tr=1.9$ (\fig{fig:spectrum}(e)) $V\simeq0.15$, for $\Gamma\tr=1.26$ (\fig{fig:phase}) $V\simeq0.31$, and for $\Gamma\tr=0.63$ (Figs.~\ref{fig:offcenter} and \ref{fig:longtime}) $V\simeq0.58$. 

The choice of $\Gamma$ is the time--frequency trade-off inherent in a physical spectrum. The filter memory $1/\Gamma$ that makes the comb visible is also the time resolution of the measurement. For $\Gamma\tr\lesssim1$ the spectrum at time $t$ averages over more than a round trip, and successive revivals begin to blur into one another, which is why the long-time spectra of \fig{fig:longtime} use the narrowest filter. 
Values $\Gamma\tr\simeq1$--$2$ are the compromise at which both the revivals and the teeth are seen, which is the point of a time-dependent spectrum.

\section{Controlling the comb: feedback phase and atomic position}
\label{sec:control}

\subsection{Feedback phase}
\label{sec:feedback}

\begin{figure}[t]
\includegraphics[width=\columnwidth]{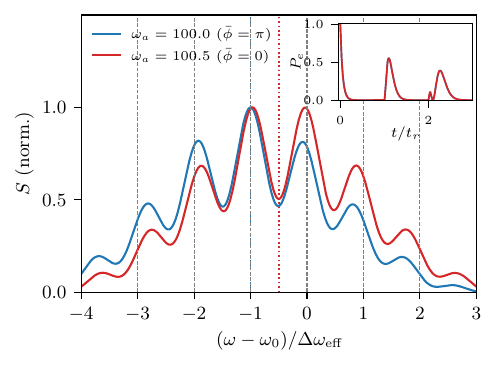}
\caption{Physical spectrum at the maximum of the first revival ($t/\tr=1.27$, $\Gamma\tr=1.26$), normalized to its maximum, for $\omega_a=100$ ($\bar\phi=\pi$) and $\omega_a=100.5$ ($\bar\phi=0$). Because the point is that the teeth stay fixed while the atomic line moves, the frequency axis is referred to a fixed $\omega_0=100$. Vertical gray dashed lines: half-cavity resonances $\omega\tr=2\pi m$; red dotted lines: the Lamb-shifted atomic frequencies $\tilde\omega_a=99.0$ and $99.5$. Inset: $P_e(t)$ for the two cases, which coincide.}
\label{fig:phase}
\end{figure}

Because the comb teeth are fixed by the cavity geometry, \eq{eq:comb_positions}, the feedback phase $\bar\phi=\tilde\omega_a\tr+\pi$ controls the position of the atomic line \emph{relative to} the comb. Shifting $\omega_a$ from $100$ to $100.5$, i.e., onto a coupled mode, changes $\bar\phi$ from $\pi$ to $0$ (equivalently, $\tilde\omega_a$ moves from $99.0$, a tooth, to $99.5$, midway between two teeth). Figure~\ref{fig:phase} shows the resulting spectra at the maximum of the first revival. For $\bar\phi=\pi$ the comb is symmetric about $\tilde\omega_a$ and has a central tooth there; for $\bar\phi=0$ the atomic line falls on a comb minimum, the spectrum is dominated by the two teeth at $\tilde\omega_a\pm\tfrac12\weff$, and the tooth positions themselves move by only $\lesssim0.1\,\weff$ through the pulling described by \eq{eq:dressed}. The populations $P_e(t)$ of the two cases are indistinguishable (inset). The physical spectrum thus resolves a phase that the population does not. For $\gamma\tr\lesssim1$ the stages overlap and the population itself becomes phase sensitive (inset of \fig{fig:compare}): for $\bar\phi\simeq0$ the feedback is constructive and the decay is accelerated, whereas for $\bar\phi\simeq\pi$ part of the excitation is trapped in a dark atom--field superposition, in close analogy with the mirror-induced suppression of decay of~\cite{Dorner2002,Tufarelli2013}.

\subsection{Influence of the atomic position}

For an atom at the position $x\ne L/2$ the two mirrors are at different distances, and radiation emitted in the two directions returns after the delays $\tau_1=2x/c$ and $\tau_2=2(L-x)/c$, with $\tau_1+\tau_2=2\tr$. Repeating the derivation of Appendix~\ref{app:dde} with $\sin^2(k_nx)=\tfrac12-\tfrac14(e^{2ik_nx}+e^{-2ik_nx})$ gives
\begin{align}
\dot{\at}(t)=&-\frac{\gamma}{2}\at(t)
-\gamma\sum_{k\ge1}e^{2ik\tilde\omega_a\tr}\,\at(t-2k\tr)\nonumber\\
&+\frac{\gamma}{2}\sum_{k\ge0}\Big[e^{i\tilde\omega_a(\tau_1+2k\tr)}\at(t-\tau_1-2k\tr)\nonumber\\
&\qquad\quad+e^{i\tilde\omega_a(\tau_2+2k\tr)}\at(t-\tau_2-2k\tr)\Big],
\label{eq:dde_offcenter}
\end{align}
where every delayed term is understood to carry a $\Theta$ function, and which reduces to \eq{eq:dde} for $\tau_1=\tau_2=\tr$. The decay rate $\gamma$ and the Lamb shift $\Delta_L=-(g_0^2/\omega_a)\sum_n\sin^2(k_nx)=-g_0^2N/2\omega_a$ are unchanged, because the spatial average of $\sin^2(k_nx)$ is $\tfrac12$ at any $x$; the envelope is therefore defined with the same $\tilde\omega_a$ as for the centered atom. The position-dependent part of the radiative shift familiar from an atom in front of a mirror \cite{Dorner2002,Cook1987} is not a constant here. It is the static limit of the delayed terms, which \eq{eq:dde_offcenter} keeps in full (Appendix~\ref{app:dde}). 

\begin{figure}[t]
\includegraphics[width=\columnwidth]{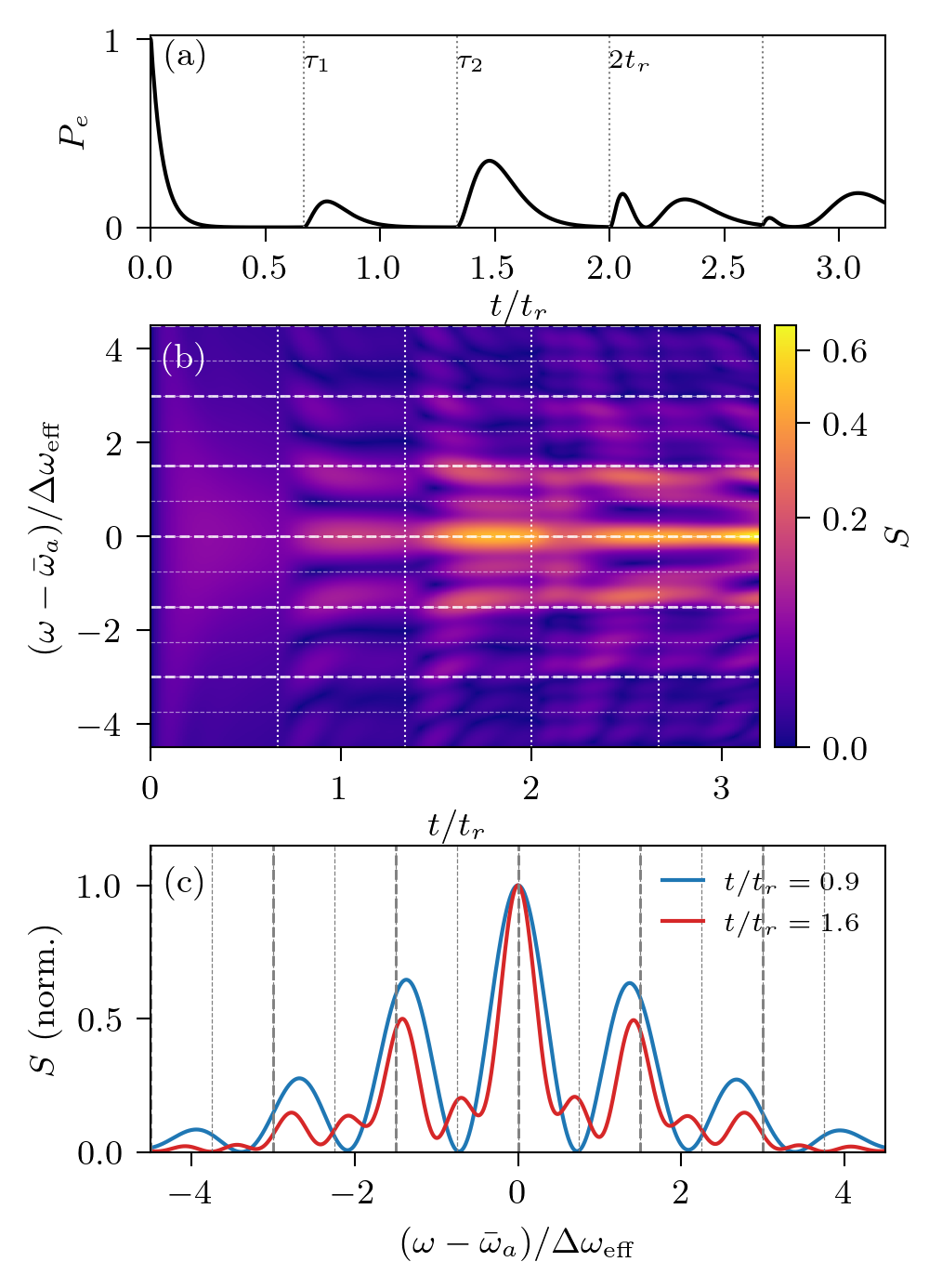}
\caption{Atom at $x=L/3$, i.e., at distances $L/3$ and $2L/3$ from the mirrors ($\tau_1=2\tr/3$, $\tau_2=4\tr/3$); other parameters as in \fig{fig:compare}, filter $\Gamma\tr=0.63$. (a) Excited-state population, with revivals at the return times $\tau_1$ (near mirror), $\tau_2$ (far mirror), $2\tr$, $\tau_1+2\tr$, and $\tau_2+2\tr$ (dotted lines). (b) Physical spectrum in the time--frequency plane (square-root color scale), with frequencies referred to $\tilde\omega_a=99$ as in all other figures. Dashed horizontal lines mark the nominal resonances of the two sub-cavities into which the reflecting atom divides the cavity, spaced by $2\pi/\tau_1=\tfrac32\weff$ (thick) and $2\pi/\tau_2=\tfrac34\weff$ (thin). (c) Normalized cuts at $t/\tr=0.9$, after the return from the near mirror only, where the spectrum shows the coarse comb of the shorter sub-cavity, and at $t/\tr=1.6$, after both returns, where the finer comb of the longer sub-cavity has appeared between its teeth; the teeth common to both sets are the strongest. As for the centered atom, the teeth are pulled slightly toward $\tilde\omega_a$ with respect to the nominal positions.}
\label{fig:offcenter}
\end{figure}

What changes is that the feedback now arrives at three interleaved sequences of times. In the half-cavity picture of Sec.~\ref{sec:combstructure} the reflecting atom now divides the cavity into two sub-cavities of lengths $x$ and $L-x$, whose resonances are spaced by $2\pi/\tau_1$ and $2\pi/\tau_2$, and each comb should appear in the spectrum when the pulse from the corresponding mirror returns. Figure~\ref{fig:offcenter} confirms this for $x=L/3$, where $2\pi/\tau_1=\tfrac32\weff$ and $2\pi/\tau_2=\tfrac34\weff$. The population [\fig{fig:offcenter}(a)] shows separate revivals at $\tau_1$, $\tau_2$, and $2\tr$, of unequal heights because the two returning pulses no longer arrive together and therefore no longer interfere constructively at the atom. The time--frequency map [\fig{fig:offcenter}(b)] and the two cuts [\fig{fig:offcenter}(c)] show the spectrum acquiring the two combs in turn: after the return from the near mirror ($\tau_1<t<\tau_2$, cut at $t/\tr=0.9$) only the coarse comb of the shorter sub-cavity is present, with teeth at $0,\pm1.4,\pm2.7$; after the return from the far mirror ($t>\tau_2$, cut at $t/\tr=1.6$) the finer comb of the longer sub-cavity appears between those teeth, at $\pm0.7,\pm2.1$. The two sets are commensurate and share every second tooth, which are correspondingly the strongest, and all teeth are pulled slightly toward $\tilde\omega_a$ from their nominal positions, as for the centered atom. The spectrum thus measures the two distances from the atom to the mirrors separately and in time order; the full round-trip delay $2\tr$, common to both sub-cavities, adds no new spacing. The special case $\tau_1\to0$ (atom near a mirror) recovers the half-cavity physics of~\cite{Dorner2002,Tufarelli2013} with a single additional delay $2\tr$. 
The spectrum thus provides a frequency-domain
signature of the atomic position through the
resonances of the two effective sub-cavities.
These features complement the population dynamics,
which already contains information about the
atom--mirror distances through the arrival times
of the returning radiation.
Together with Sec.~\ref{sec:feedback}, our results show that
the physical spectrum provides access to the
feedback phase, which is largely inaccessible
from the population dynamics in the regime
of well-separated emission stages, while
also offering a frequency-resolved characterization
of the cavity geometry.

\section{Long-time limit and dressed modes}
\label{sec:longtime}

The transient comb must eventually connect with the stationary properties of the lossless atom--cavity system. The Laplace transform of \eq{eq:dde} is (Appendix~\ref{app:dde})
\begin{equation}
\at(z)=\frac{1-q}{(z+\gamma/2)-q\,(z-\gamma/2)},\qquad q=e^{i\bar\phi}e^{-z\tr}.
\label{eq:laplace}
\end{equation}
Because the atom-cavity system is closed, the poles of $\at(z)$ lie on the imaginary axis, $z=-i\nu$, and satisfy
\begin{equation}
e^{i\omega\tr}=\frac{\gamma/2-i(\omega-\tilde\omega_a)}{\gamma/2+i(\omega-\tilde\omega_a)},\qquad \omega=\tilde\omega_a+\nu,
\label{eq:poles}
\end{equation}
i.e.,
\begin{equation}
\omega\tr=2\pi m-2\arctan\!\left[\frac{2(\omega-\tilde\omega_a)}{\gamma}\right],\qquad m\in\mathbb Z .
\label{eq:dressed}
\end{equation}
Equation~(\ref{eq:dressed}) interpolates between two limits. Far from resonance, $|\omega-\tilde\omega_a|\gg\gamma$, the arctangent tends to $\pm\pi/2$ and the roots are $\omega\tr=(2m\mp1)\pi$: the bare coupled modes $\omega=(m+\tfrac12)\weff$, since the atom is transparent there. Within the atomic linewidth, $|\omega-\tilde\omega_a|\ll\gamma$, the roots are $\omega\tr\simeq2\pi m$: the half-cavity resonances of Sec.~\ref{sec:combstructure}, since the atom acts as a mirror. Roughly $\gamma/\weff$ modes ($\simeq\pi$ for our parameters) are converted from one set to the other; these are the dressed modes of the multimode Jaynes--Cummings model \cite{Krimer2014,Sundaresan2015}. 
For $\bar{\phi}=\pi$
($\tilde{\omega}_a t_r=2\pi m$),
the atomic frequency itself is an exact root
of Eq.~\eqref{eq:dressed}. The corresponding dressed state has an atomic weight
$r_0=(1+\gamma t_r/4)^{-1}$, obtained from the residue of Eq.~\eqref{eq:laplace} at $z=0$.
For our parameters, $r_0\simeq 0.17$.
In contrast, for $\bar{\phi}=0$,
there is no dressed-mode root exactly at
$\tilde{\omega}_a$.
Equation~(\ref{eq:dressed}) reproduces the exact eigenfrequencies of the $401$-mode Hamiltonian to better than $4\times10^{-3}\,\weff$ from exact diagonalization.

\begin{figure}[t]
\includegraphics[width=\columnwidth]{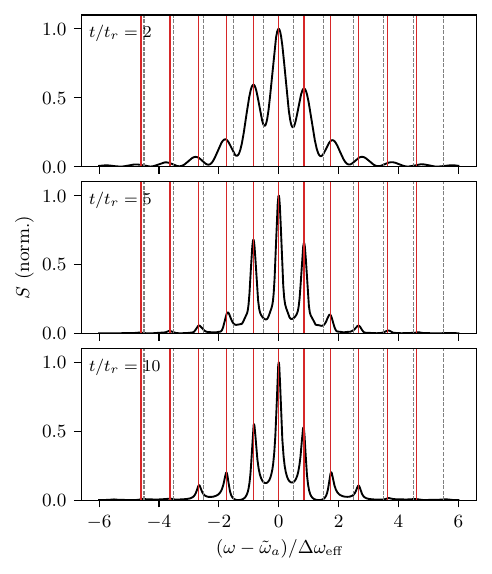}
\caption{Physical spectrum (normalized) at $t/\tr=2$, $5$, and $10$ for $\Gamma\tr=0.63$ ($\Gamma=0.1$; multimode simulation, parameters of \fig{fig:compare}). Red lines: dressed-mode frequencies from \eq{eq:dressed}; gray dashed lines: bare coupled modes, $(\omega-\tilde\omega_a)/\weff=m+\tfrac12$. Near $\tilde\omega_a$ the lines sit at the half-cavity resonances, displaced by half a mode spacing from the bare modes; far from $\tilde\omega_a$ they return to the bare modes. %
}
\label{fig:longtime}
\end{figure}

For $\Gamma t\gg1$ and $\Gamma$ smaller than the mode spacing, \eq{eq:S_alpha} reduces to $S\to2\Gamma\,|\at(z=\Gamma-i\delta)|^2$ (Appendix~\ref{app:dde}), i.e.\ to Lorentzian lines of half-width $\Gamma$ at the dressed-mode frequencies (\ref{eq:dressed}), with weights given by the residues of \eq{eq:laplace}; as $\Gamma\to0$ these become the discrete dressed-mode spectrum.
Figure~\ref{fig:longtime} shows this convergence: at $t/\tr=2$ the comb teeth are already close to the roots of \eq{eq:dressed}, and by $t/\tr=5$ and $10$ the spectrum has resolved into discrete lines at the dressed-mode frequencies, which near $\tilde\omega_a$ are displaced from the bare coupled modes by half a mode spacing.

\section{Implementation with superconducting circuits}
\label{sec:cqed}
Here we show that the Hamiltonian of \eq{eq:H} can be realized by a transmon qubit (artificial atom) coupled to a one-dimensional waveguide closed by two mirrors (Appendix~\ref{sec:circuit_qed}). The parameter that selects the regime is not the length of the waveguide but the product $\gamma\tr$ of the emission rate and the round-trip time, together with the hierarchy $\omega_a\gg\gamma,\weff$ assumed throughout. Two experimental platforms cover the two regimes discussed in this paper.

\emph{Microwave transmission lines.} For a coplanar-waveguide resonator of length $d$ and phase velocity $v\simeq1.2\times10^8\,{\rm m/s}$, $\tr=d/v$ and $\weff/2\pi=v/d$ for a centered transmon. Resonators of the usual centimeter length have $\tr\lesssim0.1$~ns and free spectral ranges of several GHz, so that the atom couples to a single mode; the multimode regime requires lengths of order a meter, which have been reached with meandered geometries \cite{Sundaresan2015}, at the price of a chip-scale device. With $d\simeq0.7$--$1$~m one has $\tr\simeq6$--$8$~ns and $\weff/2\pi\simeq120$--$170$~MHz, and with $\omega_a/2\pi\simeq5$~GHz the ratio $\omega_a/\weff\simeq30$--$40$ satisfies the hierarchy. Transmon--line couplings with $\gamma/2\pi$ of several tens of MHz are routine in waveguide QED \cite{Kannan2020}, giving $\gamma\tr\simeq1$--$3$. This is precisely the regime of the inset of \fig{fig:compare}: the emission stages overlap, the population itself depends on the feedback phase, and the comb of Sec.~\ref{sec:combstructure} is replaced by a few strongly dressed modes. Slow-wave (metamaterial) waveguides, in which the group velocity is reduced by an order of magnitude, would bring $\gamma\tr\sim10$ within reach at decimeter lengths \cite{Mirhosseini2018}. Delay-induced collapse and revival of a transmon coupled to such a waveguide has already been observed \cite{Ferreira2021PRX}. Note that circuit-QED experiments have demonstrated the simultaneous strong coupling of an artificial atom to more than 50 photonic cavity modes~\cite{Sundaresan2015}, as well as time- and frequency-resolved measurements of single-photon emission\cite{sharafiev2021visualizing}.

\emph{Surface-acoustic-wave devices.} The regime $\gamma\tr\gg1$ of Figs.~\ref{fig:compare}--\ref{fig:spectrum}(d-f), in which successive emission stages are well separated and the transient comb develops, is reached naturally when the waveguide carries phonons instead of photons. A transmon coupled through an interdigitated transducer to surface acoustic waves on a piezoelectric substrate is a two-level emitter in a one-dimensional continuum with propagation speed $v\simeq3\times10^3$~m/s, five orders of magnitude slower than light, so that a round trip of several hundred nanoseconds requires a path of a millimeter. Delays comparable to or longer than the emitter lifetime, and the resulting non-exponential decay and revivals, have been observed in this setting \cite{Andersson2019}; a transmon placed between two acoustic Bragg mirrors is precisely the model of Sec.~\ref{sec:model}, with $\tr\simeq0.1$--$1\,\mu$s, $\weff/2\pi$ of a few MHz, $\omega_a/2\pi\simeq4$~GHz, and $\gamma/2\pi$ of a few MHz, i.e., $\gamma\tr\simeq10$ or more. The microscopic derivation detailed in Appendix~\ref{sec:circuit_qed} carries over with the transmission-line modes replaced by the acoustic modes of the cavity; only the mode functions and the velocity change.

In either platform the time-dependent physical spectrum corresponds to heterodyne detection of the field leaking through a weakly coupled output port, followed by a filter of bandwidth $\Gamma$. 
Two remarks are in order. First, an output port introduces a photon-loss rate $\kappa$; the revivals survive as long as $\kappa\tr\ll1$, and the discrete lines of Sec.~\ref{sec:longtime} acquire a width $\kappa$. Second, the output field is proportional to the cavity field at the port, which is a superposition of the pulses emitted by the atom at different times, rather than to $\sigma_-(t)$ itself. Its correlation function can be computed from the same amplitudes, $G_{\rm out}(t_1,t_2)\propto\sum_{n,n'}u_n u_{n'}\beta_n^*(t_1)\beta_{n'}(t_2)$ with $u_n$ the mode functions at the port, and inserted into \eq{eq:S_def}. The comb structure and its phase dependence are inherited from the emitted field, with an additional, port-dependent spectral filter. A detailed comparison between the dipole spectrum studied here and the output spectrum is left for future work.

\section{Conclusions}
\label{sec:conclusions}

We have investigated in detail the time-dependent physical spectrum of a two-level atom coupled to a multimode cavity. By reducing the multimode problem to a delay-differential equation with an exact Laguerre-polynomial solution, Eq.~(\ref{eq:solution}), we obtained a quantitative description of both the population revivals and the spectrum. At short times the spectrum reproduces the free-space spontaneous emission. With each cavity round trip it exhibits a spectral revival, during which a transient comb of spacing $2\pi/\tr$ develops. The comb results from the quantum interference between emission stages separated by $\tr$; its teeth sit at the resonances of the two half-cavities created by the resonantly reflecting atom, its visibility scales as $e^{-\Gamma\tr}$, and at long times it converges to the dressed-mode spectrum of the atom--cavity system, which we obtained in closed form and verified against exact diagonalization. The feedback phase $\bar\phi=\tilde\omega_a\tr+\pi$ decides whether the atomic line coincides with a comb tooth or lies between two teeth, and the atomic position sets the delay structure; both are resolved by the spectrum, whereas the population revivals are insensitive to $\bar\phi$ in the regime $\gamma\tr\gg1$. 
The physical spectrum complements this information by
resolving the phase-dependent interference between successive emission stages and providing a
frequency-resolved signature of the cavity geometry.
Natural extensions include driven atoms, where the comb should coexist with the Mollow triplet~\cite{Ancheyta_PRA_2016}; several emitters sharing the same delay lines; the output spectrum through a realistic port; and giant artificial atoms \cite{Andersson2019,Kannan2020}, which couple to the waveguide at several points separated by a propagation time and therefore obey a delay-differential equation of the same form as~\eq{eq:dde_offcenter}, with the delays set by the coupling-point separations rather than by mirrors. Finally, for a giant artificial atom, the corresponding time-dependent physical spectrum should resolve interference between emission from the different coupling points in the same way it resolves the successive reflections in this work. All of these can be addressed with the formalism we developed here.
\begin{acknowledgments}
This work  has been funded by SECIHTI--Mexico, under grant No. CBF2023-2024-2888, and by DGAPA-PAPIIT-UNAM under grant No. IA104625. R.R.-A. thanks DGAPA-UNAM, Mexico for support under Project No. IA101826. J.R. thanks DGAPA-UNAM Project IN 117925.
D. A.-L. would like to express his gratitude to SECIHTI--Mexico for his PhD Scholarship.
\end{acknowledgments}
\appendix
\section{Atomic correlation function}
\label{app:corr}

The first-order correlation function is $G(t_1,t_2)=\langle\sigma_+(t_1)\sigma_-(t_2)\rangle$. For $t_1\ge t_2$,
\begin{align}
G(t_1,t_2)&=\langle\psi(0)|U^\dagger(t_1)\sigma_+U(t_1)U^\dagger(t_2)\sigma_-U(t_2)|\psi(0)\rangle\nonumber\\
&=\langle\psi(t_1)|\sigma_+U(t_1-t_2)\sigma_-|\psi(t_2)\rangle,
\end{align}
with $U(t)=e^{-iHt/\hbar}$. Acting with the ladder operators on the single-excitation state (\ref{eq:state}) gives $\sigma_-|\psi(t_2)\rangle=\alpha(t_2)|g,\vec0\rangle$ and $\langle\psi(t_1)|\sigma_+=\alpha^*(t_1)\langle g,\vec0|$, so that
\begin{equation}
G(t_1,t_2)=\alpha^*(t_1)\alpha(t_2)\langle g,\vec0|U(t_1-t_2)|g,\vec0\rangle .
\end{equation}
Since $|g,\vec0\rangle$ is an eigenstate of $H$ with eigenvalue $0$ (our choice of energy zero), $U(t)|g,\vec0\rangle=|g,\vec0\rangle$, and
\begin{equation}
G(t_1,t_2)=\alpha^*(t_1)\alpha(t_2),
\end{equation}
which is \eq{eq:G_alpha}; the case $t_1<t_2$ follows from $G(t_2,t_1)=G^*(t_1,t_2)$. Substituting into \eq{eq:S_def}, the double integral factorizes into $|\int_0^t dt'\,e^{(\Gamma+i\omega)t'}\alpha(t')|^2$, and inserting $\alpha=e^{-i\tilde\omega_a t}\at$ gives the second form of \eq{eq:S_alpha} with $\delta=\omega-\tilde\omega_a$.

\section{Delay-differential equation, Lamb shift, and exact solution}
\label{app:dde}

\paragraph*{Exact equation in the Laplace domain.} Integrating \eq{eq:beta} with $\beta_n(0)=0$ and substituting into \eq{eq:alpha} gives the integro-differential equation (\ref{eq:integro}). Its Laplace transform, with $\alpha(0)=1$ and $\alpha(z)=\int_0^\infty e^{-zt}\alpha(t)\,dt$, is
\begin{equation}
\left[z+i\omega_a+\Sigma(z)\right]\alpha(z)=1,\quad 
\Sigma(z)=\sum_n\frac{g_n^2}{z+i\omega_n},
\label{eq:selfenergy}
\end{equation}
where $\Sigma(z)$ is the Laplace transform of the kernel $K(\tau)$. The envelope (\ref{eq:envelope}) has the transform $\at(z)=\alpha(z-i\tilde\omega_a)$, so
\begin{equation}\label{eq:selfenergy_env}
\begin{split}
    1=&\left[z-i\Delta_L+\Sigma(z-i\tilde\omega_a)\right]\at(z),\\
\Sigma(z-i\tilde\omega_a)=&\sum_n\frac{g_n^2}{z+i(\omega_n-\tilde\omega_a)}
\end{split}
\end{equation}
with $\Delta_L\equiv\tilde\omega_a-\omega_a$ still to be determined. Everything that follows consists of evaluating this sum for the coupling (\ref{eq:gn}). For a centered atom, $g_n^2=g_0^2(\omega_n/\omega_a)\sin^2(n\pi/2)$ selects the odd modes, $\omega_n=(m+\tfrac12)\weff$ with $m$ integer and $\weff=2\pi/\tr$, and we write, identically,
\begin{equation}
g_n^2=g_0^2+g_0^2\,\frac{\omega_n-\omega_a}{\omega_a},
\qquad
\Sigma=\Sigma_{\rm flat}+\Sigma_1 .
\label{eq:split}
\end{equation}

\paragraph*{Linear part: the Lamb shift.} Using $(\omega_n-\omega_a)/[z+i(\omega_n-\tilde\omega_a)]=-i+i(z-i\Delta_L)/[z+i(\omega_n-\tilde\omega_a)]$, the contribution of the linear term is, exactly,
\begin{equation}
\Sigma_1(z-i\tilde\omega_a)=-i\,\frac{g_0^2N_c}{\omega_a}+\frac{i(z-i\Delta_L)}{\omega_a}\,\Sigma_{\rm flat}(z-i\tilde\omega_a),
\label{eq:Sigma1}
\end{equation}
with $N_c$ the number of coupled modes. The first term is a constant, purely imaginary self-energy, i.e., a frequency shift of the atomic line. On the imaginary axis, $z=-i\nu$, the second term is $(\nu+\Delta_L)/\omega_a$ times $\Sigma_{\rm flat}$, smaller than $\Sigma_{\rm flat}$ by $\sim10^{-2}$ for all frequencies of interest, and we drop it. We now \emph{choose} $\Delta_L$ so that the constant cancels the explicit $-i\Delta_L$ in \eq{eq:selfenergy_env}:
\begin{equation}
\Delta_L=-\frac{g_0^2N_c}{\omega_a}=-1.
\end{equation}
This is the full principal-value sum of \eq{eq:lamb} because its flat part, $g_0^2\sum_n(\omega_a-\omega_n)^{-1}$, vanishes by pairwise cancellation for a mode spectrum symmetric about $\omega_a$, as in our simulations ($0<\omega_n<2\omega_a$). In general the flat part of the principal-value sum is \emph{not} to be added to $\Delta_L$: it is the static ($\nu\to0$) limit of $\Sigma_{\rm flat}$, which the delay-differential equation retains exactly through its delayed terms, and absorbing it into $\tilde\omega_a$ would count it twice. (For an off-center atom, Sec.~\ref{sec:control}, this flat part is nonzero and is the position-dependent radiative shift of an atom near a mirror; it is carried by the delayed terms of \eq{eq:dde_offcenter}.) $\Delta_L$ grows linearly with the cutoff, as is usual for the Lamb shift in rotating-wave models, and is absorbed into the renormalized frequency $\tilde\omega_a=\omega_a+\Delta_L$. With these identifications \eq{eq:selfenergy_env} reduces to
\begin{equation}
\left[z+\Sigma_{\rm flat}(z-i\tilde\omega_a)\right]\at(z)=1 .
\label{eq:Sigma_approx}
\end{equation}

\paragraph*{Flat part: the Dirac comb.} Extending the sum over $m$ to all integers, which neglects the modes far from resonance and the cutoff (corrections of relative order $\gamma/\omega_a$ and $\weff/\omega_a$), the Poisson summation formula $\sum_m e^{-im\weff\tau}=\tr\sum_k\delta(\tau-k\tr)$ gives the time-domain kernel of the envelope equation,
\begin{equation}
\begin{split}
\bar K(\tau)&=g_0^2\sum_{m}e^{-i[(m+\frac12)\weff-\tilde\omega_a]\tau}
\\
&=\gamma\sum_{k\ge0}e^{ik(\tilde\omega_a\tr+\pi)}\,\delta(\tau-k\tr),
\label{eq:Kflat}
\end{split}
\end{equation}
with $\gamma=g_0^2\tr=2\pi g_0^2/\weff$ and the delta function at $\tau=0$ counted with weight $\tfrac12$ because it sits at the boundary of the integration range; the factor $e^{-ik\pi}=(-1)^k$ is the reflection phase accumulated after $k$ round trips. This is \eq{eq:kernel}, with the feedback phase $\bar\phi=\tilde\omega_a\tr+\pi$ of \eq{eq:phibar} appearing because the comb is evaluated at the renormalized frequency; its time-domain form is the delay-differential equation (\ref{eq:dde}). In the Laplace domain,
\begin{equation}
\Sigma_{\rm flat}(z-i\tilde\omega_a)=\gamma\Big[\frac12+\sum_{k\ge1}q^k\Big]
=\frac{\gamma}{2}\,\frac{1+q}{1-q},
\label{eq:Sigmaflat}
\end{equation}
with $q=e^{i\bar\phi}e^{-z\tr}$, for ${\rm Re}\,z>0$, and \eq{eq:Sigma_approx} gives
\begin{equation}
\Big[z+\frac{\gamma}{2}\,\frac{1+q}{1-q}\Big]\at(z)=1,
\end{equation}
which, solved for $\at(z)$, is \eq{eq:laplace}.

\paragraph*{Inverse Laplace transform.} For ${\rm Re}\,z>0$, $|q|<1$ and
\begin{equation}
\at(z)=(1-q)\sum_{m\ge0}q^m\frac{(z-\gamma/2)^m}{(z+\gamma/2)^{m+1}} .
\end{equation}
The inverse transform of $(z-\gamma/2)^m/(z+\gamma/2)^{m+1}$ is $e^{-\gamma t/2}L_m(\gamma t)$, as follows from the shift $z\to z-\gamma/2$ and from $\mathcal L^{-1}[(z-\gamma)^m/z^{m+1}]=\sum_j\binom{m}{j}(-\gamma t)^j/j!=L_m(\gamma t)$. The factor $q^m=e^{im\bar\phi}e^{-zm\tr}$ delays this function by $m\tr$ and multiplies it by $e^{im\bar\phi}$; the factor $(1-q)$ subtracts the same function delayed by one further round trip. Collecting the terms with the same delay $k\tr$ gives
\begin{equation}
\at(t)=\sum_{k\ge0}e^{ik\bar\phi}\,e^{-\gamma s_k/2}\left[L_k(\gamma s_k)-L_{k-1}(\gamma s_k)\right]\Theta(s_k),
\end{equation}
with $s_k=t-k\tr$, which is \eq{eq:solution}. The identity $L_k-L_{k-1}=L_k^{(-1)}=-(y/k)L^{(1)}_{k-1}$ gives \eq{eq:terms}. The Lamb shift enters the envelope only through $\bar\phi$; in the amplitude $\alpha(t)=e^{-i\tilde\omega_at}\at(t)$ it also appears as the Wigner--Weisskopf shift of the carrier.

\paragraph*{Poles.} The poles of $\at(z)$ satisfy $(z+\gamma/2)=q\,(z-\gamma/2)$. Setting $z=-i\nu$ gives $e^{i\bar\phi}e^{i\nu\tr}=-(\gamma/2-i\nu)/(\gamma/2+i\nu)$, which with $\bar\phi=\tilde\omega_a\tr+\pi$ and $\omega=\tilde\omega_a+\nu$ is \eq{eq:poles}; writing the right-hand side as $e^{-2i\arctan(2\nu/\gamma)}$ gives \eq{eq:dressed}. Since the right-hand side has unit modulus, all roots are real: the closed atom--cavity system has a purely discrete spectrum. The agreement between these roots and the exact eigenfrequencies of the multimode Hamiltonian quoted in Sec.~\ref{sec:longtime} (to $0.004\,\weff$) is a direct check of the approximations made in (\ref{eq:Sigma_approx}) and (\ref{eq:Kflat}).

\paragraph*{Long-time spectrum.} Writing the filter weight inside the integral, \eq{eq:S_alpha} reads $S=2\Gamma|\int_0^t dt'\,e^{-\Gamma(t-t')}e^{i\delta t'}\at(t')|^2$. Expanding the envelope in the dressed modes, $\at(t')=\sum_j r_je^{-i\nu_jt'}$, with $\nu_j$ the roots of \eq{eq:dressed} and $r_j$ the residues of \eq{eq:laplace}, the integral gives for $\Gamma t\gg1$
\begin{eqnarray}
S\to2\Gamma\Big|\sum_j\frac{r_j\,e^{i(\delta-\nu_j)t}}{\Gamma+i(\delta-\nu_j)}\Big|^2
&\simeq&2\Gamma\sum_j\frac{|r_j|^2}{\Gamma^2+(\delta-\nu_j)^2}\nonumber\\
&=&2\Gamma\,|\at(z=\Gamma-i\delta)|^2, \nonumber
\label{eq:S_longtime}
\end{eqnarray}
where the cross terms between different modes oscillate in $t$ and average out when $\Gamma$ is smaller than the mode spacing. The spectrum is thus a set of Lorentzian lines of half-width $\Gamma$ centered at the dressed-mode frequencies and weighted by $|r_j|^2$, the overlap of the initial state with each dressed state; since $2\Gamma/(\Gamma^2+x^2)\to2\pi\delta(x)$, it becomes the discrete dressed-mode spectrum as $\Gamma\to0$. The filter bandwidth is what regularizes the Laplace transform on the imaginary axis, which for the closed system does not converge by itself.

\section{Visibility of the comb}
\label{app:visibility}

\paragraph*{Two-stage contrast.} During the $k$th revival, $k\tr\le t<(k+1)\tr$, the stages $j<k-1$ in \eq{eq:S_comb} are suppressed relative to the $k$th by $e^{-(k-j)\Gamma\tr}$ and by the shorter collection time of the latest stage, so we keep the two most recent terms. Writing $\theta\equiv\bar\phi+\delta\tr$,
\begin{align}
S&\simeq2\Gamma e^{-2\Gamma t}e^{2k\Gamma\tr}\,\big|F_k+e^{-i\theta}e^{-\Gamma\tr}F_{k-1}\big|^2\nonumber\\
&=2\Gamma e^{-2\Gamma t}e^{2k\Gamma\tr}|F_k|^2\big[1+|c_k|^2+2|c_k|\cos(\theta-\arg c_k)\big],
\label{eq:twobeam}
\end{align}
with $c_k=e^{-\Gamma\tr}F_{k-1}/F_k$ as in \eq{eq:visibility}. 
Equation~(\ref{eq:twobeam}) is a two-beam interference pattern in the variable $\theta$, which advances by $2\pi$ when $\delta$ advances by $\weff=2\pi/\tr$. The single-stage amplitudes $F_j(\delta)$ are Fourier transforms of envelopes of duration $\gtrsim1/\gamma$ and therefore vary on the scale $\gamma$ in $\delta$; for $\gamma\gg\weff$ they, and hence $c_k$, are nearly constant across one tooth. Under this condition the extrema of \eq{eq:twobeam} occur at $\cos(\theta-\arg c_k)=\pm1$,
\begin{equation}
S_{\max,\min}^{}\propto|F_k|^2\,(1\pm|c_k|)^2,
\end{equation}
and the Michelson contrast is \cite{Mandel_Wolf_1995}
\begin{equation}
V_k=\frac{S_{\max}-S_{\min}}{S_{\max}+S_{\min}}=\frac{2|c_k|}{1+|c_k|^2},
\label{eq:michelson}
\end{equation}
which is \eq{eq:visibility}. For $|c_k|\ll1$, $V_k\simeq2|c_k|=2e^{-\Gamma\tr}|F_{k-1}/F_k|$; the full form (\ref{eq:michelson}) is bounded by 1 and must be used when $\Gamma\tr\lesssim1$. The maxima of (\ref{eq:twobeam}) lie at $\theta=\arg c_k+2\pi m$. For $k=1$ at $\delta=0$, $F_0$ and $F_1$ are real with $F_0>0$ and $F_1<0$ [see \eq{eq:Fk} and $L_1^{(-1)}(y)=-y$], so $\arg c_1=\pi$ and, with $\theta=\omega\tr+\pi$, the maxima are at $\omega\tr=2\pi m$, which is \eq{eq:comb_positions}.

\paragraph*{Single-stage amplitudes.} From \eq{eq:Fk}, with $p\equiv\gamma/2-\Gamma-i\delta$ and $T_j=t-j\tr$,
\begin{align}
F_0(\delta;T_0)&=\frac{1-e^{-pT_0}}{p},\\
F_1(\delta;T_1)&=-\gamma\,\frac{1-e^{-pT_1}(1+pT_1)}{p^2}.
\end{align}
For $\Gamma<\gamma/2$ and $T_j\gg1/\gamma$ these tend to $F_0\to1/p$ and $F_1\to-\gamma/p^2$, so that at $\delta=0$
\begin{equation}
|c_1|\to e^{-\Gamma\tr}\Big(\frac12-\frac{\Gamma}{\gamma}\Big),\qquad
V_1\simeq e^{-\Gamma\tr}\Big(1-\frac{2\Gamma}{\gamma}\Big)
\label{eq:V1_asym}
\end{equation}
for small contrast. The dashed and dotted curves of \fig{fig:visibility} are \eq{eq:michelson} with the finite-time and the asymptotic $c_1$, respectively; at $t/\tr=1.27$ and $\Gamma\tr=1.26$ the first stage has been fully collected ($pT_0\simeq11$) but the second only partly ($pT_1\simeq2.4$), which is why the two curves differ and why the finite-time one lies above.

\paragraph*{Numerical definition.} The visibility extracted from the computed spectra in \fig{fig:visibility} is the mean of $(S_{\max}-S_{\min})/(S_{\max}+S_{\min})$ over each local maximum of $S(\omega)$ and its nearest local minimum within $|\delta|<1.5\,\weff$, i.e., over the central three teeth, where the envelope is flattest and the signal largest. Taking the global extrema instead would measure the envelope rather than the modulation. The definition is biased by the slope of the envelope within a tooth and by the variation of $c_k$ across the central window, where $|F_0/F_1|$ is smaller than at $\delta=0$; these effects, together with the partial collection of the second stage, account for the $\sim20\%$ spread between the measured points and the two analytic curves in \fig{fig:visibility}. When the modulation falls below the resolution of the extremum search ($\Gamma\tr\ge4.4$) the procedure returns zero.

\section{Circuit-QED model: transmon coupled to a multimode transmission-line resonator}
\label{sec:circuit_qed}

\paragraph*{Resonator modes.} We consider a transmission-line resonator of length $L$ short-circuited at both ends, with inductance $l_0$ and capacitance $c_0$ per unit length and phase velocity $v=1/\sqrt{l_0c_0}$. Its Hamiltonian is
\begin{equation}
H_r=\int_0^L dx\left[\frac{Q^2(x)}{2c_0}+\frac{1}{2l_0}\Big(\frac{\partial\Phi(x)}{\partial x}\Big)^2\right],
\label{eq:Hr}
\end{equation}
with $\Phi(x)$ the node flux and $Q(x)$ the charge per unit length. The boundary conditions $\Phi(0)=\Phi(L)=0$ give the normal modes
\begin{align}
\Phi(x)&=\sum_{n=1}^{\infty}u_n(x)\,\Phi_n,\qquad u_n(x)=\sqrt2\sin(k_nx),
\nonumber\\
k_n&=\frac{n\pi}{L},\qquad \omega_n=vk_n,
\label{eq:modes}
\end{align}
normalized as $L^{-1}\int_0^Lu_nu_{n'}\,dx=\delta_{nn'}$; these are the mode frequencies of Sec.~\ref{sec:model} with $c\to v$. Each mode is a harmonic oscillator of capacitance $C_r=c_0L$ and impedance $Z_n=1/C_r\omega_n$. With
\begin{equation}
\Phi_n=\sqrt{\frac{\hbar Z_n}{2}}\,(a_n+a_n^\dagger),\quad
Q_n=i\sqrt{\frac{\hbar}{2Z_n}}\,(a_n^\dagger-a_n),
\end{equation}
one has $H_r=\sum_n\hbar\omega_na_n^\dagger a_n$ up to the zero-point energy. The voltage at a point $x$ is
\begin{equation}
V(x)=\frac{Q(x)}{c_0}=i\sum_n\sqrt{\frac{\hbar\omega_n}{c_0L}}\,\sin(k_nx)\,(a_n^\dagger-a_n),
\label{eq:Vx}
\end{equation}
whose prefactor $\sqrt{\hbar\omega_n/c_0L}$ is the zero-point voltage at an antinode. It vanishes at the ends, which therefore act as the mirrors of Sec.~\ref{sec:model}.

\paragraph*{Transmon and coupling.} A transmon \cite{Koch2007} of charging energy $E_C=e^2/2C_\Sigma$ and Josephson energy $E_J$, with charge $q=2en$ and phase $\varphi$, is coupled capacitively through $C_g$ at the position $x$. The circuit Hamiltonian is
\begin{align}
H&=\frac{[\,q-C_gV(x)\,]^2}{2C_\Sigma}-E_J\cos\varphi+H_r
\nonumber\\
&=H_{\rm tr}+H_r-\frac{C_g}{C_\Sigma}\,q\,V(x)+\frac{C_g^2}{2C_\Sigma}\,V^2(x),
\label{eq:Hcircuit}
\end{align}
with $H_{\rm tr}=4E_Cn^2-E_J\cos\varphi$. The last term renormalizes the resonator capacitance locally and is dropped in the weak-coupling limit $C_g\ll C_\Sigma$. Restricting the transmon to its two lowest levels $|g\rangle$ and $|e\rangle$, with transition frequency $\omega_a$ and charge matrix element $q_{eg}=|\langle g|q|e\rangle|$, and dropping a constant,
\begin{equation}
H_{\rm tr}\simeq\hbar\omega_a\sigma_+\sigma_-,\qquad q\simeq q_{eg}(\sigma_++\sigma_-).
\end{equation}
This truncation requires the anharmonicity $E_C/\hbar$ to exceed the emission bandwidth $\gamma$. Inserting \eq{eq:Vx} into \eq{eq:Hcircuit} and keeping the excitation-conserving terms (rotating-wave approximation), results in
\begin{align}
H\simeq\;&\hbar\omega_a\sigma_+\sigma_-+\sum_n\hbar\omega_na_n^\dagger a_n
\nonumber\\
&-\sum_n\hbar g_n(x)\left(a_n\sigma_++a_n^\dagger\sigma_-\right),
\label{eq:HJC}
\end{align}
where the phase of the mode operators has been redefined, $a_n\to-ia_n$, to make the coupling real, and
\begin{align}
\hbar g_n(x)&=\frac{C_g}{C_\Sigma}\,q_{eg}\sqrt{\frac{\hbar\omega_n}{c_0L}}\,\sin(k_nx)
\nonumber\\
&=\hbar g_0\sqrt{\frac{\omega_n}{\omega_a}}\,\sin(k_nx),\qquad
g_0=\frac{C_g}{C_\Sigma}\,q_{eg}\sqrt{\frac{\omega_a}{\hbar c_0L}} .
\label{eq:g0circuit}
\end{align}
Equations (\ref{eq:HJC}) and (\ref{eq:g0circuit}) are exactly the model of Eqs.~(\ref{eq:H}) and (\ref{eq:gn}), with the transition dipole $d_{eg}/\sqrt{\epsilon_0}$ replaced by the effective charge $(C_g/C_\Sigma)q_{eg}/\sqrt{c_0}$. The $\sqrt{\omega_n}$ dependence of the coupling, which produces the Lamb shift of Sec.~\ref{sec:dde}, is the zero-point voltage of each mode, and a transmon at the center of the line couples only to the odd modes, as in the main text. The cutoff frequency $\omega_{\rm cut}$ is set physically by the frequency above which the lumped description of the coupling fails, of the order of the transmon plasma frequency.

\bibliography{refs}
\end{document}